\documentclass[fleqn,10pt]{wlscirep}
\usepackage[utf8]{inputenc}
\usepackage[T1]{fontenc}
\usepackage{multirow} 
\usepackage{appendix}
\usepackage{lscape}
\usepackage{comment}

\title{Multilayer Network of Cardiovascular Diseases and Depression via Multipartite Projection}

\author[1,*]{Jie Li}
\author[1]{Cillian Hourican}
\author[4, 5, 6]{Pashupati P. Mishra}
\author[4, 5, 6]{Binisha H. Mishra}
\author[5, 7]{Mika K\"ah\"onen}
\author[8, 9, 10,11]{Olli T. Raitakari}
\author[5, 12]{Reijo Laaksonen}
\author[13, 14, 15]{Mika Ala-Korpela}
\author[16]{Liisa Keltikangas-J\"arvinen}
\author[17,18]{Markus Juonala}
\author[4, 5, 6]{Terho Lehtim\"aki}
\author[2]{Jos A. Bosch}
\author[1,3]{Rick Quax}

\affil[1]{Computational Science Lab, Informatics Institute, University of Amsterdam, Amsterdam, The Netherlands}
\affil[2]{Clinical Psychology, Faculty of Social and Behavioural Sciences, University of Amsterdam, Amsterdam, The Netherlands}
\affil[3]{Institute for Advanced Study, Amsterdam, The Netherlands}
\affil[4]{Department of Clinical Chemistry, Faculty of Medicine and Health Technology, Tampere University, Tampere, Finland}
\affil[5]{Finnish Cardiovascular Research Center Tampere, Faculty of Medicine and Health Technology, Tampere University, Tampere, Finland}
\affil[6]{Department of Clinical Chemistry, Fimlab Laboratories, Tampere, Finland}
\affil[7]{Department of Clinical Physiology, Tampere University Hospital, Tampere, Finland}
\affil[8]{Research Centre of Applied and Preventive Cardiovascular Medicine, University of Turku, Turku, Finland}
\affil[9]{Department of Clinical Physiology and Nuclear Medicine, Turku University Hospital, Turku, Finland}
\affil[10]{Centre for Population Health Research, University of Turku and Turku University Hospital, Turku, Finland}
\affil[11]{InFLAMES Research Flagship, University of Turku, Turku, Finland}
\affil[12]{Zora Biosciences Oy, Espoo, Finland}
\affil[13]{Systems Epidemiology, Faculty of Medicine, University of Oulu and Biocenter Oulu, Oulu, Finland}
\affil[14]{Research Unit of Population Health, Faculty of Medicine, University of Oulu, Oulu, Finland}
\affil[15]{NMR Metabolomics Laboratory, School of Pharmacy, University of Eastern Finland, Kuopio, Finland}
\affil[16]{Department of Psychology and Logopedics, University of Helsinki, Helsinki, Finland}
\affil[17]{Division of Medicine, Turku University Hospital, Turku, Finland}
\affil[18]{Department of Medicine, University of Turku, Turku, Finland}

\affil[*]{j.li4@uva.nl}


\begin{abstract}

Cardiovascular diseases (CVD) and depression exhibit significant comorbidity, which is highly predictive of poor clinical outcomes. Yet, the underlying biological pathways remain challenging to decipher, presumably due to the non-linear associations across multiple mechanisms. In this study, we introduced a multipartite projection method based on mutual information correlations to construct multilayer disease networks as a novel approach to explore such intricate relationships. We applied this method to a cross-sectional dataset from a wave of the Young Finns Study, which includes data on CVD and depression, along with related risk factors and two omics of biomarkers: metabolites and lipids. Rather than directly correlating CVD-related phenotypes and depressive symptoms, we extended the notion of bipartite networks to create a multipartite network, linking these phenotypes and symptoms to intermediate biological variables. Projecting from these intermediate variables results in a weighted multilayer network, where each link between CVD and depression variables is marked by its ‘layer’ (i.e., metabolome or lipidome). Applying this projection method, we identified potential mediating biomarkers that connect CVD with depression. These included metabolites such as creatinine, valine, phospholipids in very large high-density lipoproteins (HDL), triglycerides in small low-density lipoproteins (LDL), and apolipoprotein B; as well as lipids such as specific sphingomyelins (SM), phosphatidylcholines (PC), triacylglycerols (TAG) and diacylglycerols (DAG). These biomarkers may therefore play significant roles in the biological pathways underlying CVD-depression comorbidity. Additionally, the network projection highlighted sex and BMI as key risk factors, or confounders, in this comorbidity. Our method is scalable to incorporate any number of omics layers and various disease phenotypes, offering a comprehensive, system-level perspective on the biological pathways contributing to comorbidity.

\end{abstract}
\begin{document}

\flushbottom
\maketitle
%
%
\thispagestyle{empty}


\section*{Introduction}


Cardiovascular diseases (CVD) and depression present major global health challenges with a significant co-occurrence yielding profound comorbidity~\cite{Alexander2007, Hert2018, Mishra2024}. Indeed, individuals with depression face an elevated risk of developing CVD~\cite{Katherine2023}. Conversely, individuals with CVD are more prone to developing depression, and their comorbidity is linked to a 3-fold higher mortality rate~\cite{BAUNE2012, Hare2013}. Thus, where each phenotype alone already significantly diminishes quality of life, imposes a considerable burden on the healthcare system, and remains challenging to treat \textemdash their combination presents an even bigger challenge. In either direction, both behavioral and biological factors play a role~\cite{BAUNE2012, Dhar2016, KAHL2019}. An extensive literature explored possible biological mechanisms underlying this bidirectional relationship, such as increased inflammation, endocrine disturbances, and imbalances in neurotransmitter levels, as well as dysregulation of the sympathetic and parasympathetic nervous systems~\cite{Halaris2017, Shao2020, WU2021, lee2023}. However, the precise nature of these interactions has not yet clearly been established and continued efforts to understand their biological associations remain important. In this article, we contribute to this effort by introducing a novel method for constructing disease networks, aimed at elucidating the relationships between CVD phenotypes, depressive symptoms, and their biological mediators.

The concept of ’human disease networks’  has emerged as a powerful framework to study human diseases and underlying biological mechanisms using a network perspective~\cite{Goh2007}. As an illustrative example, Goh et al. introduced a projection method to reconstruct disease networks based on a bipartite graph~\cite{Goh2007}. In their projected networks, the association between two diseases is established if they share a gene implicated in both conditions. The strength of gene-disease associations was quantified by (linear) Pearson correlation, and the strength of disease-disease associations (i.e., their comorbidity) was defined as the number of genes that were significantly associated with both diseases. Although an important innovation, this methodology presents several limitations that might lead to biases in estimating disease associations. Firstly, the method used only a single-omics layer (i.e., genes) to connect diseases, whereas in reality there are multiple omics layers (e.g., genes, proteins, lipids, etc.) connecting diseases. Secondly, simply counting the number of shared intermediate biomarkers is a rather coarse measure that weighs each biomarker equally, irrespective of the the strength of their (Pearson) correlations. Thirdly, the use of Pearson correlation overlooks nonlinear correlations~\cite{Justin2014}. Lastly, the proposed method does not allow incorporation of higher-order correlations into the projection.

Several studies have sought to address these biases by expanding on the original methodology. For instance, Park et al. implemented cellular networks, including genes and proteins, into their analyses to provide a multi-omics view~\cite{Park2009}. To tackle the second limitation, Zhou et al. applied the cosine similarity of disease symptom vectors to quantify disease associations, and defined interaction orders based on the path length between proteins~\cite{Zhou2014}. Taking a different approach, Grosdidier et al. proposed a concept of molecular comorbidity index (MCI) based on proteins common between two diseases to estimate the strength of the association~\cite{Grosdidier2014}. While these advancements extended the scope to multiple omics or refined the measurement of disease associations, significant methodological gaps remain. Specifically, no existing method provides comprehensive integrative solution that incorporates multiple omics datasets into a single network analysis, employs a more refined quantification of mediating biomarkers, and leverages non-linear correlations for richer expressiveness.

To innovate on existing methodologies and resolve the above gaps, we introduced a multipartite projection method that advances the study of disease networks by: (\textit{i}) incorporating any number of omics datasets and various diseases; (\textit{ii}) utilizing the non-parametric Shannon mutual information (MI) correlation; and (\textit{iii}) employing a weighted approach to count shared intermediate biomarkers and risk factors. MI is a non-parametric metric that captures both linear and nonlinear mutual interdependencies between variables~\cite{Alexander2023}, offering a more comprehensive measure of association than conventional methods. By counting biomarkers in a weighted manner, our method additionally provides a finer-grained analyses to quantify the links between diseases. We proposed three increasingly complex and refined definitions for quantifying projected disease links. These definitions not only consider the number of shared biomarkers, but also account for the strength of their nonlinear correlations, and the proximity of their neighboring biomarkers. This approach allows for a more comprehensive understanding of potential shared etiologies compared to direct correlation methods.

We applied the proposed projection method to a dataset from the Young Finns Study (YFS) and selected a projection definition that demonstrated the most effective performance. The dataset encompasses three primary sets of variables: (1) CVD and depression phenotypes (see Table~\ref{tab:1} and ~\ref{tab:2}), (2) related risk factors (Table~\ref{tab:importance}), (3) intermediate omics biomarkers (metabolites and lipids). Our method and its resulting weighted multilayer networks allow us to: (\textit{i}) identify key biomarkers across multiple biological layers that contribute to CVD-depression comorbidity; (\textit{ii}) assess the relative importance of risk factors by integrating them into the projected network; (\textit{iii}) identify key biomarkers related to a particular projected link between disease phenotypes or between a risk factor and a disease phenotype. The projection method identified sex and body mass index (BMI) as two important risk factors related to both CVD and depression, and uncovered significant mediating biomarkers underlying their comorbidity. These biomarkers included metabolites such as creatinine, valine, phospholipids in very large high-density lipoproteins (HDL), triglycerides in small low-density lipoproteins (LDL), and apolipoprotein B, as well as lipids such as specific sphingomyelins (SM), phosphatidylcholines (PC), triacylglycerols (TAG) and diacylglycerols (DAG). By accommodating multiple omics datasets and considering non-linear correlations, our method provides a comprehensive, system-level view on the biological pathways and mechanisms contributing to comorbidities. This approach offers a valuable tool for future research, facilitating a deeper understanding of complex disease interactions.

\section*{Methods}

\subsection*{The Cardiovascular Risk in Young Finns Study}

Our proposed method was applied to the 2007 wave of the Young Finns Study (YFS)~\cite{Raitakari2008}. This cross-sectional follow-up study provides a multidimensional dataset that encompasses: (1) CVD-related phenotypes, depressive symptoms, (2) related risk factors (i.e., covariates), (3) intermediate omics biomarkers (metabolites, lipids). In CVD-related variables, the ideal cardiovascular health (CVH) score is a composite measure that typically includes factors such as blood pressure, cholesterol levels, diet, physical activity, and smoking status. Higher scores reflect better health outcomes across these metrics, suggesting a lower risk of cardiovascular disease. The Beck Depression Inventory (BDI) score is a summary variable derived from individual depressive symptoms. A higher BDI score indicates worse mental health, as it reflects a higher level of depressive symptoms. This multidimensional dataset enables the identification of significant biomarkers, and the inclusion of two omics datasets facilitates the use of our multipartite projection approach.

For each participant, the phenotype variables consist of 17 CVD-related indicators and six risk factors (see Table ~\ref{tab:1} and ~\ref{tab:importance}). Depressive variables include the BDI score and 21 individual depressive symptoms contributing to the calculation of the BDI score (see Table~\ref{tab:2}). The two omics datasets comprise 228 metabolomic variables and 437 lipidomic variables, which were stored in two separate tables. Detailed information about these variables and the methods used for their measurement can be found in Section S1.1 of the Supplementary Information.

\subsection*{Missing data imputation and data discretization}

We used random sample imputation to replace each missing value of a variable with a random sample drawn from existing values of that variable. This conservative approach minimizes the risk of introducing spurious associations, with the impact of imputation diminishing as the proportion of missing values increases. After imputation, we retained the data points in which all omics data were available, resulting in a final sample of 1686 participants (983 women, 703 men). Continuous variables were discretized by using a quantile-based method, dividing each continuous variable into probabilistically equal-sized buckets. Discretization was needed because part of the data was discrete and information-theoretic metrics require comparable data types. The optimal number of quantiles was determined by Sturges’ rule. Categorical and discrete variables were included without modification.


\subsection*{Significant MI correlation network}
Mutual information (MI) was calculated on the discretized data to quantify the correlation between the YFS variables. This information-theoretic approach allows for capturing both linear and nonlinear relationships. It measures the average amount of information communicated in one random variable $X$ about another $Y$ ~\cite{Alexander2023}.
\begin{equation}
MI(X;Y) = \sum_{y \in Y}\sum_{x \in X} p(x,y)\cdot log \frac{p(x,y)}{p(x)p(y)}, \label{eq:MI}
\end{equation}
where $x$ and $y$ are the observations of variables $X$ and $Y$, $p(x)$ and $p(y)$ are marginal probabilities, $p(x,y)$ is the joint probability mass function. 

Before constructing the significant MI network, we first applied a filtering process to minimize double counting and reduce redundant information among variables. The step was necessary because our proposed method incorporates the strength of MI correlations into the projection, and ensuring that each correlation is uniquely represented without redundancy is crucial for the method's effectiveness. Consequently, some repeated variables and one variable from each highly correlated pair based on MI correlation were excluded from the analysis (see Section S1.1 of the Supplementary Information). These variables included the BDI score, CVD-related variables (such as "fmd40", "fmd40pr", "fmd60", "fmd60pr", "fmd80", "fmd80pr", "maxfmd", "imtmax", "bbmax" and "volscore"), and another 115 variables in metabolites and lipids. This process resulted in a final dataset of 584 variables for further analysis.

We employed the bootstrapping method to compute $p$-values for MI correlations. Only correlations with $p$-values smaller than the significance level ($\alpha=0.01$) were included in the network, resulting in a significant multipartite MI correlation network. The corresponding adjacency matrix of the significant correlation network is denoted as $F(X;Y)$, in which each element is $f(X;Y)$. 


\subsection*{Multipartite projection method}

The MI can quantify correlations directly between variables of phenotypes, symptoms and risk factors. However, these direct correlations would not be able to suggest any plausible mechanisms (pathways) between them, let alone identify key biomarkers. Our multipartite projection method reconstructs these disease links indirectly as projected scores, based on their MI correlations with intermediate biomarkers.

We provided three definitions for the projected score. In the main analysis, we selected the most effective definition based on a comparative analysis (see Section S2.1 of the Supplementary Information). This definition calculates the projected score as the sum of the average MI correlations between each pair of variables and their shared neighboring nodes (intermediate biomarkers). It is formulated as follows:
\begin{equation}
w(X_i;X_j) = \sum_{k} \frac{f(X_i;Y_k)+f(X_j;Y_k)}{2}, \label{eq:proj}
\end{equation}
Where $X_i$ and $X_j$ are risk factor and/or phenotype variables and $Y_k$ is an intermediate biomarker that is a shared neighboring node of $X_i$ and $X_j$ in the tripartite MI network. Both $f(X_i;Y_k)$ and $f(X_j;Y_k)$ are non-zero. 

\begin{figure}[ht]
\centering
\includegraphics[width=\linewidth]{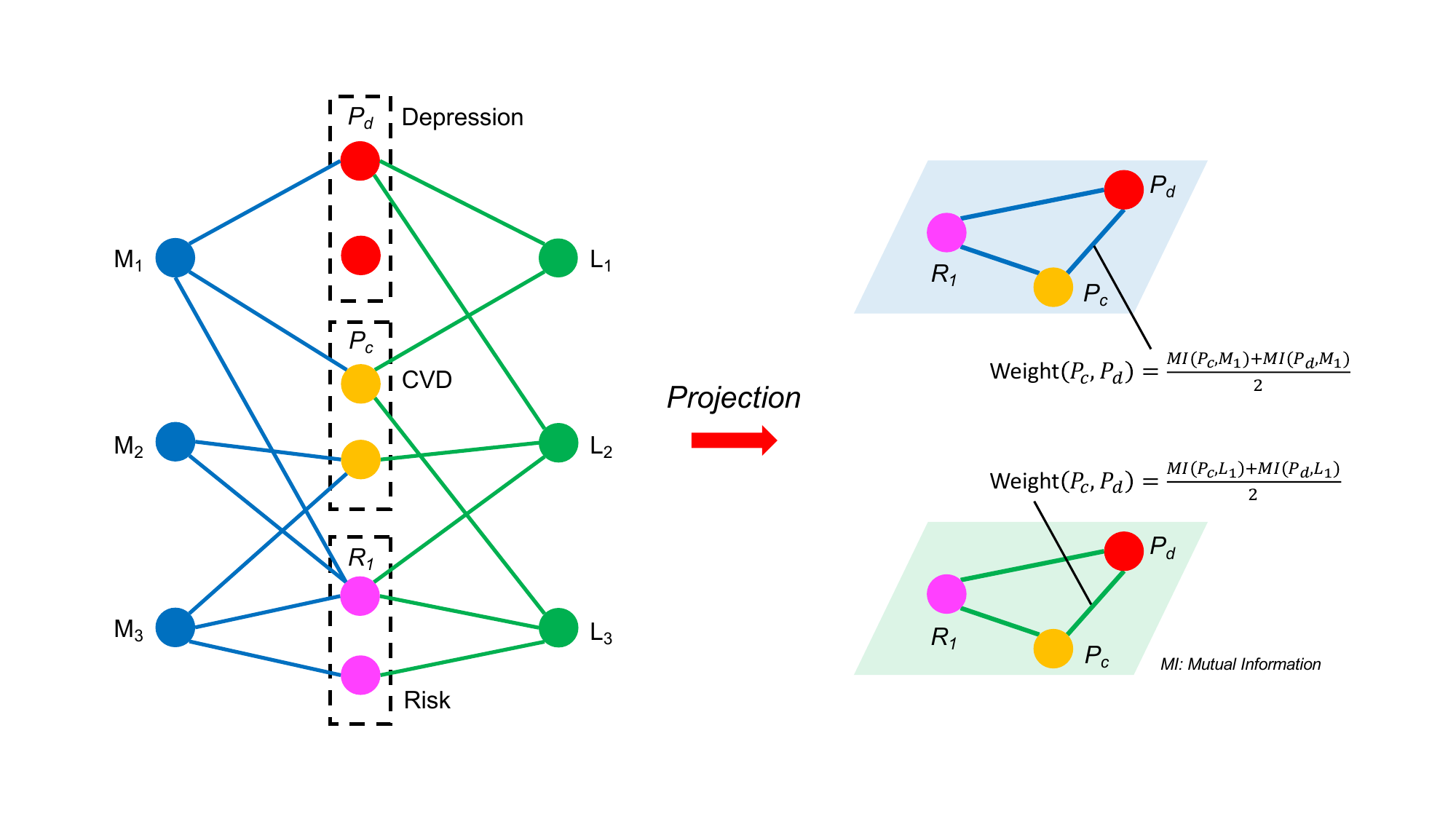}
\caption{\textbf{Stylized description of projection method used in the main analysis}. The network on the left is a tripartite network, in which blue nodes depict metabolomic variables, green nodes represent lipidomic variables, and nodes in the middle include red nodes representing depressive symptoms, yellow nodes representing CVD-related phenotypes, purple nodes representing risk factors. The figures on the right show the projected multilayer networks on blue and green panels. The blue one shows the metabolomic layer of the projected network. The green one presents the lipidomic layer of the projected network.}
\label{fig:stylized}
\end{figure}

This projection definition considers intermediate biomarkers in a weighted manner. When projected to different types of biomarkers, this leads to a weighted, multilayer disease network of CVD phenotypes, depressive symptoms and related risk factors, in which each layer of the network corresponds to the type of biomarker involved in the projection. A stylized representation of this projection definition using average MI correlation is depicted in Figure~\ref{fig:stylized}.

\subsection*{Contribution of biomarkers to projected scores}

The projection method enables us to calculate the contribution of each biomarker (in our case, metabolites and lipids) to the projected score of each link between diseases. Considering the projection definition used in the main analysis and our particular focus on CVD-depression comorbidity, we defined the total contribution score of a biomarker to all possible projected links between CVD phenotypes and depressive symptoms as follows: 
\begin{equation}
CON(Y_k)= \sum_{i,j} \frac{f(X_i;Y_k)+f(X_j;Y_k)}{2} \label{eq:ttl}
\end{equation}
where $Y_k$ is a biomarker, $X_i$ and $X_j$ represent pairs of phenotypes and symptoms. Both $f(X_i;Y_k)$ and $f(X_j;Y_k)$ are non-zero. Such a contribution score enables the identification of significant biomarkers that contribute the most to the overall comorbidity between CVD and depression. For a given pair of $(X_i, X_j)$, the method identifies the significant biomarkers that contribute the most to the projected score of the specific link between these diseases.

\subsection*{The relative importance of risk factors related to CVD and depression}

It is widely recognized that CVD and depression are often linked to behavioral and demographic risk factors~\cite{Chaplin23, lee2023}. Therefore, risk factors were included in the projection. This approach allows for the assessment of their relative importance concerning specific phenotypes. The relative importance of a risk factor with respect to a particular phenotype is defined as the ratio of the projected score between the phenotype and the risk factor to the total projected score between the phenotype and all risk factors. This can be formulated as:
\begin{equation}
r(Z_i, X_j)=\frac{w(Z_i;X_j)}{\sum_{i} w(Z_i;X_j)} \label{eq:importance}
\end{equation}
where $X_j$ is a phenotype and $Z_i$ is one of the risk factors that has a link to the phenotype. $w(Z_i;X_j)$ is the projected score between the phenotype and the risk factor. The relative importance of risk factors considering CVD and depression is computed as the average of $r(Z_i, X_j)$ over all CVD-related phenotype and depressive symptoms, respectively. For a particular strong projected link between a risk factor and a disease, we can also identify the significant biomarkers contributing to the projected score using Equation ~\ref{eq:ttl}.

\section*{Results}

\subsection*{Significant tripartite MI network and projected multilayer disease networks}

\begin{figure}[ht]
\centering
\includegraphics[width=0.8\linewidth]{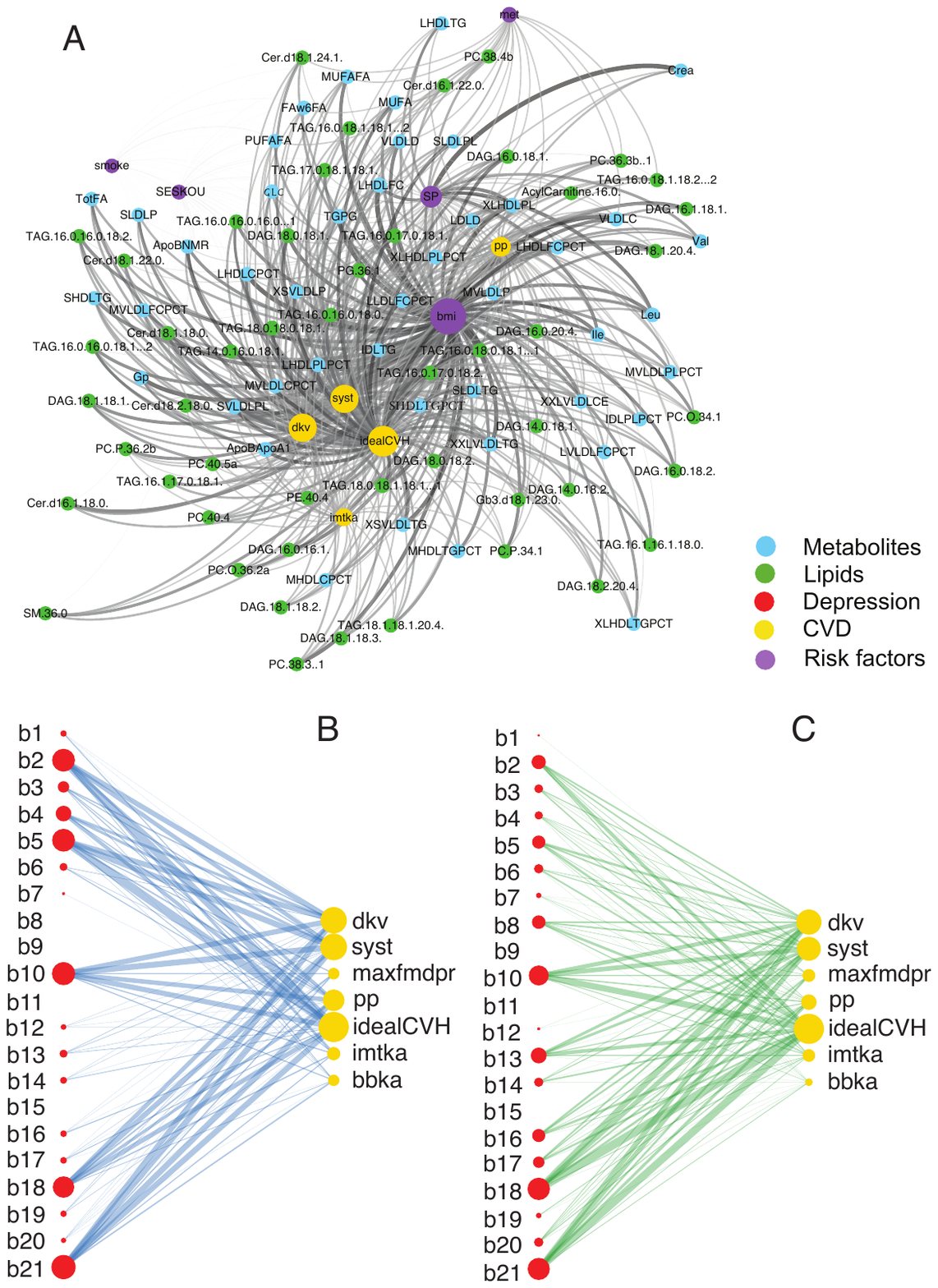}
\caption{\textbf{The multipartite projection and weighted projected multilayer disease networks.} \textbf{A}: The significant tripartite MI correlation network ($p-value<0.01$), in which variables are partitioned into three groups: metabolites, lipids, and phenotypes (including CVD-related phenotypes, depressive symptoms and related risk factors). \textbf{B}: The metabolomic layer of the projected bipartite network of CVD-related phenotypes and depressive symptoms. \textbf{C}: The lipidomic layer of the projected bipartite network.}
\label{fig:projection}
\end{figure}

The projection method builds on the construction of a significant multipartite MI correlation network. As shown in Figure~\ref{fig:projection}A, this network is tripartite, in which variables are partitioned into three groups: metabolome (blue), lipidome (green), and the combined group of phenotypes/symptoms of CVD (yellow) and depression (red) along with risk factors (purple). It shows only the links between metabolites/lipids and variables in the combined group. The edge width is determined by the mean MI value, which was calculated over 20 iterations of random sample imputation. Node size is proportional to the node’s weighted degree, reflecting the strengths of its connections. To enhance readability and visualization, nodes in the tripartite network were filtered based on their weighted degree, retaining 105 nodes and 669 edges. Consequently, no depressive symptom variables are visible in this network due to their weak correlations with intermediate biomarkers. However, it is important to note that this filtering process was applied only for visualization purposes; the projection analysis itself was performed on the unfiltered network, and depressive symptoms were thus included in the analysis.

Within the tripartite network, certain risk factors such as sex and BMI, CVD-related phenotypes like the ideal CVH score, average diastolic, and systolic blood pressure, exhibit a high weighted degree, indicating extensive correlations with intermediate biomarkers. However, interpreting these correlations based solely on this network poses challenges, as it remains unclear which specific biomarkers are driving these strong associations. This limitation underscores the difficulty in elucidating the underlying causative mechanisms and biological pathways from the correlation network alone.

Figure~\ref{fig:projection}B and C illustrate the metabolomic layer and lipidomic layer of the projected network of CVD-related phenotypes and depressive symptoms, respectively. In these projected networks, the link (edge) width is quantified by the mean projected score over 20 runs of random sample imputation. The node size is proportional to its weighted degree. We only include the projected links between CVD and depression variables due to our focus on the comorbidity between these two conditions. The projected networks are thus displayed in a bipartite layout. In the metabolomic layer of the projected network, we find strong projected links between five individual depressive symptoms (pessimism ('b2'), guilty feelings ('b5'), crying ('b10'), changes in appetite ('b18'), loss of libido ('b21')) and three CVD-related variables (average diastolic blood pressure ('dkv'), average systolic blood pressure ('syst'), and ideal CVH score ('idealCVH')). In the lipidomic layer, strong projected links are observed between three depressive symptoms ('b10', 'b18' and 'b21') and the same three CVD-related phenotypes. These strong projected links suggest a high risk of co-occurrence between these phenotypes and symptoms, predicting multimorbid phenotypes of CVD and depression.

\begin{figure}[ht]
\centering
\includegraphics[width=0.6\linewidth]{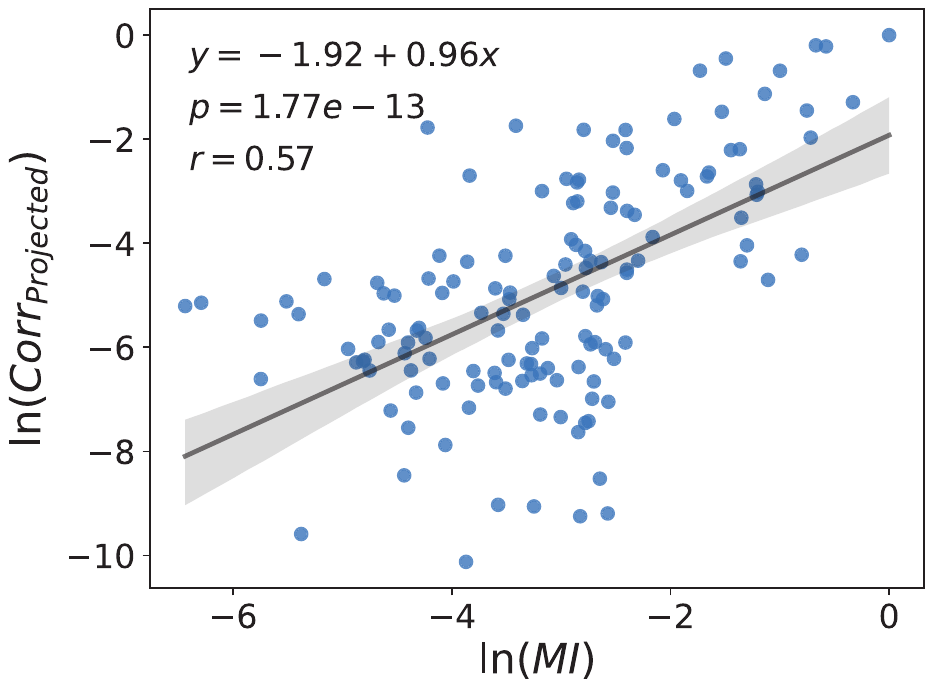}
\caption{\textbf{Projected score vs. MI correlation on a log-log scale.} The logarithmic projected score and MI correlation are fitted by a linear model. $p$ is the p-value of the slope. $r$ is their Pearson correlation coefficient. This plot shows a roughly linear relationship between the MI correlation and projected score on a log-log scale, confirming the rationality of the multipartite projection method.}
\label{fig:proj_vs_mi}
\end{figure}

Figure~\ref{fig:proj_vs_mi} compares the values of the MI correlation between CVD and depression with their projected score values from the projection method, showing a roughly positive linear relationship on a log-log scale. This pattern suggests that the projection method is effective, yet there remains a substantial degree of variability. We anticipate that this variability arises from the fact that the analysis includes only two omics of biomarkers. We expect that including additional categories of biomarkers would improve the correlation further. 

\subsection*{Mediating biomarkers related to the comorbidity between CVD and depression}

To investigate the biological pathways underlying CVD-depression comorbidities, we calculated the mean total contribution score for each metabolite and lipid across all projected links between CVD-related phenotypes and depressive symptoms over 20 runs of random imputation. Figure~\ref{fig:ttl_cont}A and B display the top ten mediating metabolites and lipids that contribute the most to these projected CVD-depression links. The top mediating metabolites include creatinine, valine, leucine, phospholipids in very large HDL and small VLDL, triglycerides in small LDL and very small VLDL, and apolipoprotein B. The top mediating lipids consist of several specific sphingomyelins (SM), phosphatidylcholines (PC), triacylglycerols (TAG), diacylglycerols (DAG), phosphatidylglycerols (PG), and phosphatidylethanolamines (PE). These metabolites and lipids were identified as significant biomarkers that link CVD and depression through biological pathways such as inflammation, cholesterol regulation, and atherosclerosis. Most of these metabolites and some lipids are already known, while others, such as SM and PG, are potentially novel biomarkers. Detailed rankings and further information about these biomarkers can be found in Table~\ref{tab:ttl_imp}. 


\subsection*{Relative importance of risk factors and related biomarkers}

\begin{figure}[ht]
\centering
\includegraphics[width=0.8\linewidth]{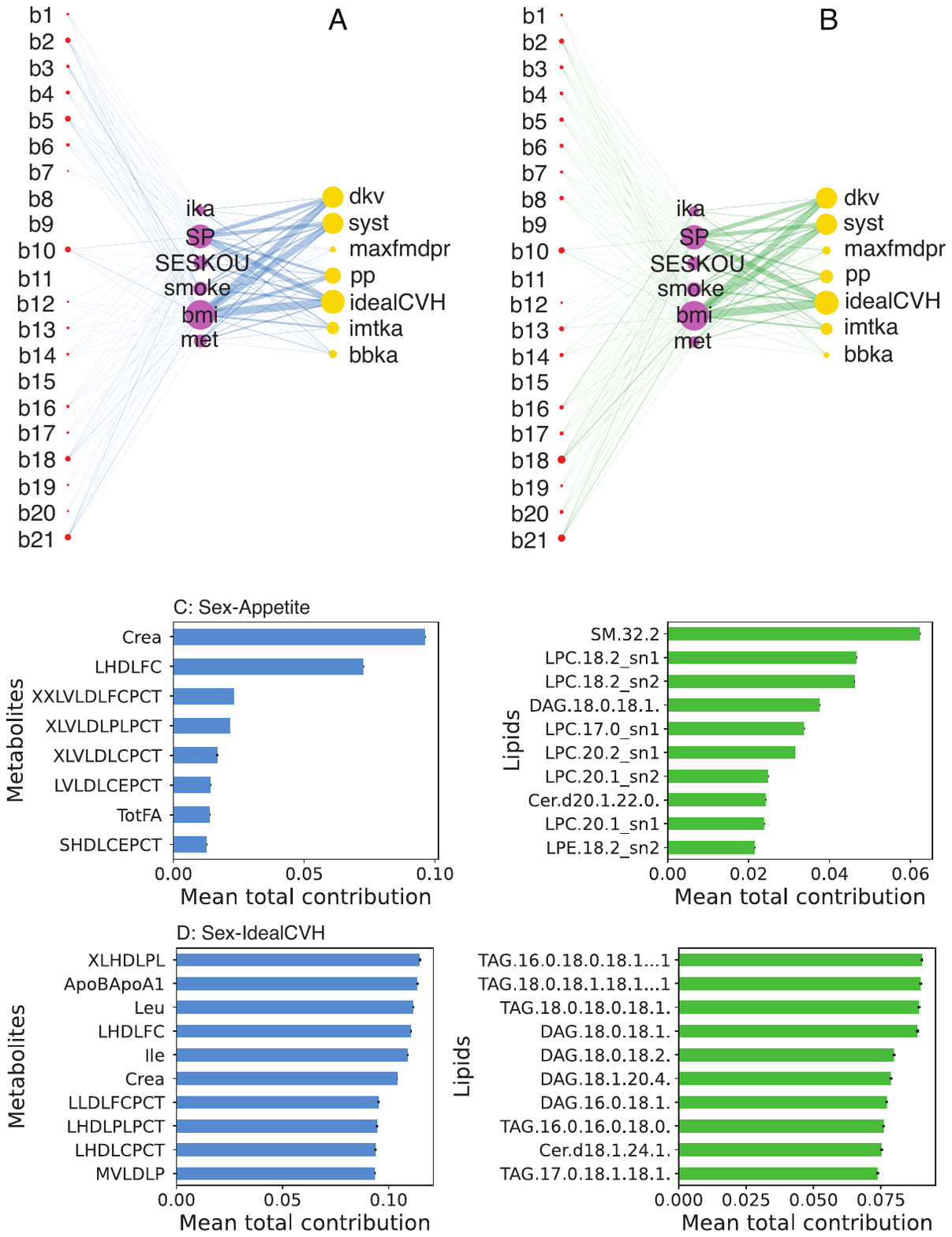}
\caption{\textbf{Projected multilayer network of CVD phenotypes, depressive symptoms, and related risk factors; and the top ten mediating biomarkers related to specific risk-phenotype links.} \textbf{A}: The metabolomic layer of the projected network. \textbf{B}: The lipidomic layer of the projected network. Yellow, red, and purple nodes represent CVD phenotypes, depressive symptoms, and risk factors, respectively. \textbf{C}: The top ten mediating metabolites and lipids contributing to the strong projected link between sex ('SP') and changes in appetite ('b18'). Only eight metabolites are in the ranking of metabolites because there are less than 10 metabolites contributing to the link. \textbf{D}: The top 10 mediating metabolites and lipids contributing to the strong projected link between sex ('SP') and the ideal cardiovascular health score ('IdealCVH').}
\label{fig:risk_proj}
\end{figure}

Including risk factors in the projection results in a projected multilayer network, in which nodes are partitioned into three groups: risk factors, CVD-related phenotypes, and depressive symptoms (see Figure~\ref{fig:risk_proj}A and B). Observing both metabolomic and lipidomic layers of the projected network, the high weighted degree of sex and BMI indicates their extensive associations with both CVD and depression compared to other risk factors. The relative importance of all risk factors is shown in Table~\ref{tab:importance}. The results suggest that sex and BMI are two important risk factors related to both CVD and depression, which corroborates previous studies~\cite{Marco2000, mosca2011, Noh2015, khan2018}. It is important to note that sex can be a covariate or confounder. As a result, we conducted gender-specific analyses (see Section S2.3 of the Supplementary Information). Additionally, age--another well-known risk factor--does not show high importance in the YFS dataset, likely because the YFS participants were all young adults.

We selected two demonstrated high-score projected links ("Sex"--"Appetite ('b18')" and "Sex"--"idealCVH") to further analyze the contributing biomarkers. In this case, sex may act as a confounding factor, influencing both variables. Due to this confounding effect, it is plausible that appetite and the ideal CVH score show a strong projected score in the projected multilayer networks (see Figure~\ref{fig:projection} B and C). The significant mediating biomarkers based on the mean total contribution score are shown in Figure~\ref{fig:risk_proj}C, D. Details of these biomarkers can be found in Table \textcolor{blue}{S6} and \textcolor{blue}{S7}. For the link between sex and appetite, significant metabolites include creatinine, free cholesterol in large HDL, and phospholipids, cholesterol and triglycerides in very large VLDL (very low-density lipoprotein). Significant lipids are mainly sphingomyelins (SM) 32:2 and lysophosphatidylcholines (LPC). For the link between sex and the ideal CVH score, significant metabolites encompass phospholipids, free and total cholesterol in large HDL and LDL, apolipoprotein B, and Leucine. Significant lipids are primarily triacylglycerols (TAG) and diacylglycerols (DAG). These identified biomarkers shed light on how risk factors influence the development of CVD and depression.

\section*{Discussion}

In this study, we proposed a multipartite projection method to infer complex disease interactions between CVD and depression, resulting in the creation of weighted multilayer networks encompassing CVD-related phenotypes and depressive symptoms. The proposed method can be considered a generalization of previous disease network modeling approaches~\cite{Goh2007, Park2009, Zhou2014}. It demonstrates several key advantages by employing mutual information (MI) correlation and incorporating multi-omics datasets. First, MI captures both linear and nonlinear associations, providing a more comprehensive understanding of the interdependencies between variables. Second, incorporating multiple omics datasets into a single projection allows for a richer and more detailed analysis of the biological pathways involved. Third, the method introduces a refined projection strategy that accounts for shared biomarkers in a weighted manner, thereby offering a more nuanced approach to quantifying disease associations. Moreover, the proposed method extends beyond existing approaches by offering the potential to incorporate directed causal relationships and higher-order interactions, such as synergistic associations, into the projection. This capability is particularly important for understanding the complex and often non-linear interactions that may underlie comorbid conditions like CVD and depression. It also has the capability to construct multilayer higher-order disease networks by considering the shared neighboring nodes of more than two disease variables.

We applied the method to the YFS dataset~\cite{Raitakari2008} and inferred the projected scores between CVD-related phenotypes and depressive symptoms considering two available omics datasets: metabolome and lipidome. A weighted multilayer disease network was created based on the projected scores. Each layer of the projected network represents the links between diseases in a specific omics context, which is metabolomic or lipidomic in this study. This framework allows for a rich network representation of CVD and depression across multiple dimensions and facilitates the identification of related biomarkers. It provides better insights into possible biological pathways and mechanisms underlying the projected links between these diseases, getting closer to explaining comorbidities.

\begin{figure}[ht]
\centering
\includegraphics[width=\linewidth]{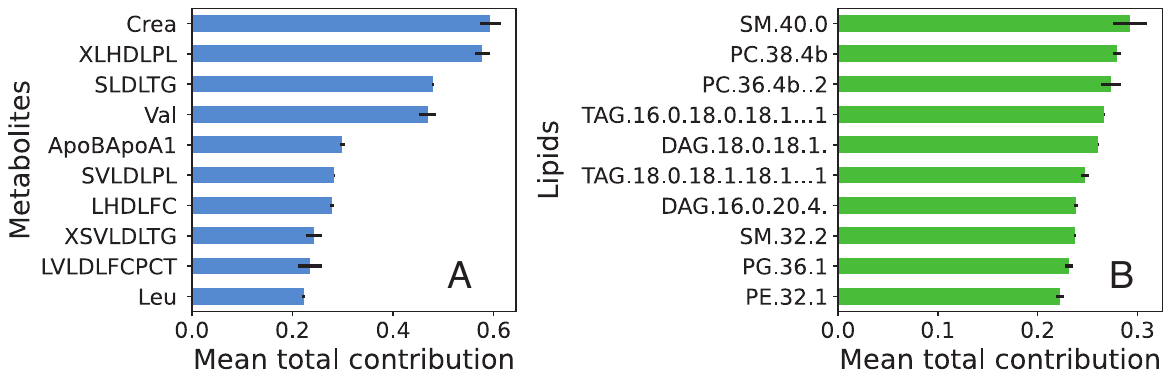}
\caption{\textbf{The top 10 significant mediating metabolites and lipids related to CVD-depression comorbidity (all links between CVD phenotypes and depressive symptoms).} \textbf{A}: The top ten significant metabolites in terms of mean total contribution score over 20 random imputations. \textbf{B}: The top ten significant lipids in terms of mean total contribution score over 20 random imputations. X-axis is the mean total contribution score and the error bar represents the unbiased standard error of the mean score, over 20 random imputations.}
\label{fig:ttl_cont}
\end{figure}

The metabolites and lipids identified as the most significant contributors (sorted in Figure~\ref{fig:ttl_cont}) are inferred to play key roles in the biological pathways connecting CVD and depression. Among metabolites, creatinine (a byproduct of creatine metabolism) emerged as a significant factor. In the context of depression, creatine has antidepressant-like effects~\cite{pazini2019possible, bakian2020dietary}. Conversely, in the realm of CVD, elevated creatinine levels are associated with traditional cardiovascular risk factors such as diabetes, dyslipidemia in hypertensive patients, and stroke~\cite{Chen23}. These findings can also be observed in our study, as detailed in Section S2.2 of the Supplementary Information. Similarly, valine, a branched-chain amino acid (BCAA), has been highlighted in recent studies for its significant association with depression~\cite{Whipp2022, Gammoh2024}. Elevated levels of BCAAs, including both valine and leucine, have been linked to an increased risk of developing CVD~\cite{Whipp2022, hu2023causal}. These particles are also rich in phospholipids, which are known to be instrumental in protecting against cardiovascular disease by promoting cholesterol efflux and reducing inflammation~\cite{Nicholls2005, Kontush2014}.

Phospholipids within high-density lipoprotein (HDL) particles are particularly noteworthy for their anti-inflammatory effects, which may not only protect against CVD but also improve mood and enhance the efficacy of conventional antidepressant treatments in patients with major depression~\cite{Glaser2015}. This anti-inflammatory property is also associated with apolipoprotein A-I, a major protein component of HDL particles~\cite{Tomio2013}, which may explain the high ranking of variable "ApoBApoA1" in our analysis. Further, high cholesterol has long been recognized as one of the significant risk factors for CVD~\cite{jeong2018}. It also shows a positive correlation with depression severity~\cite{wagner2019}.

The projection method also identified several specific lipids, including sphingomyelins (SM), phosphatidylcholines (PC), triacylglycerols (TAG), diacylglycerols (DAG), phosphatidylglycerols (PG), and phosphatidylethanolamines (PE), as significant biomarkers (Figure~\ref{fig:ttl_cont}). Among these, PC and PE are two major classes of phospholipids that are associated with CVD and depression. They play crucial roles in various biological processes, including those related to CVD risk and outcomes~\cite{Miao2024}, and are the most abundant phospholipids in mammalian cell membranes. Alterations in PC and PE, have been observed in individuals with psychotic experiences, which are associated with an increased risk of mental disorders, including depression~\cite{YIN2022}. Additionally, TAG and DAG have also been reported to be associated with both CVD and depression in many studies~\cite{Michael11, bot2020, sellem2023, Charilaos2024}.

Notably, the projection method identified SM 32:2, SM 40:0 and PG 36:1, as significant biomarkers associated with both CVD and depression. While direct evidence linking these biomarkers to CVD and depression is limited, exploring novel biomarkers like SM and PG in the research of CVD-depressioin comorbidity could offer new insights into the underlying biological pathways. Currently, SM 32:2 (SM(d18:2/14:0)) shows a significant and positive association with atherosclerosis, a key risk factor for CVD~\cite{sojo2023plasma}. Additionally, some SM species have been reported to be linked with depression and anxiety~\cite{demirkan2013plasma, Walther18}. PG is an intermediate in the biosynthesis of various lipids, particularly cardiolipin (CL)~\cite{SCHERER2011, morita2015}, which may be associated with mitochondrial function and cardiac health~\cite{Zheni2015}. However, further research is needed to elucidate the precise role of SM and PG in the comorbidity between CVD and depression.

When turning our attention to risk factors, the projected network identified sex and BMI as two important risk factors linking CVD and depression. Sex is indeed a risk factor (e.g., covariate, confounder) for these health conditions~\cite{Georgios2017, lee2023}. Men, on average, are more prone to certain risks, such as high blood pressure, higher levels of LDL (bad) cholesterol, and a higher likelihood of smoking and excessive alcohol consumption. Consequently, the risk of developing CVD tend to begin earlier in men, often around middle age. Conversely, women are more likely to be diagnosed with depression~\cite{whiteford2013global}. The sex-specific difference in depression can be attributed to the sex-induced biological changes such as in hormones in women, and different presentations of depressive symptoms~\cite{albert2015}. It is important to note that sex can act as a covariate or confounder. Thus, we conducted gender-specific analyses using the projection method. Further discussion on the results is provided in Section S2.3 of the Supplementary Information. 

BMI is considered to be highly associated with both CVD and depression. High BMI, particularly when it falls into the obese category (BMI of 30 or higher), is a well-established risk factor for cardiovascular diseases~\cite{Bastien2014, Francisco2016}. Depression shows a significant U-shaped association with BMI categories (underweight, normal, overweight and obesity)~\cite{de2009depression, badillo2022}. In addition, age is another important risk factor associated with both CVD and depression~\cite{Mirowsky1992}, but not in the YFS as the participants in this study were young adults.

The biomarkers underlying the strong links between risk factors and phenotypes were also identified. These biomarkers are considered important in elucidating how risk factors contribute to disease development and understanding the underlying biological pathways involved. For instance, creatinine is a gender-specific metabolite~\cite{Leary17, Sakon24}, and has been recognized as a nutritional marker associated with decreased appetite~\cite{lopes2007, Claudia15}. SM 32.2 and many specific LPCs are identified as biomarkers related to appetite in individuals with depression. Although no human studies have reported this finding, a recent paper using mouse models discovered that an increase in LPCs in the blood following a prolonged fasting increased lysophosphatidic acid (LPA) level in the brain~\cite{endle2022}. These LPAs can modulate the excitability of nerve cells in the cerebral cortex, thereby influencing food intake and appetite regulation. Additionally, the detection of TAG and DAG in the context of sex-related cardiovascular research suggests potential lipid-related mechanisms underlying sex disparities in CVD susceptibility and outcomes~\cite{Tabassum2022}. TAG concentrations are consistently lower in healthy lean and obese women than in men~\cite{zhang2019Age}, and its elevated level has been reported to be highly associated with CVD~\cite{Hokanson1998, sarwar2007, Michael11}.

This study has several limitations. Firstly, we conducted two steps to eliminate redundant information based on MI correlation. However, this process might lead to an underestimation of projected scores, as some intermediate biomarkers that are not entirely redundant may have been omitted. Secondly, the YFS dataset includes only limited omics data: metabolome and lipidome, which may not be sufficient for a complete construction of the projected links. These two limitations likely contribute to the consistently low explaining score (see Section S2.1 of the Supplementary Information) for all projection definitions. This finding underscores the importance of including multiple omics data in the projection. Another limitation pertains to the absence of directionality in the MI correlation. MI is a correlation measure rather than a causal discovery method, which makes it incapable of capturing causal relationships between variables. It is therefore challenging to determine whether a biomarker or risk factor plays a role of a confounder, collider, mediator or moderator. Finally, our method is blind to so-called higher-order correlations or synergistic associations beyond pairwise relationships. This is a common limitation in the field, as similar approaches in the literature have the same constraint.


This study may provide direction for future data collection efforts towards biomarkers that are unexpectedly predicted as highly important, such as specific SM and PG. Future work should also involve incorporating directed causal~\cite{Jhee2019} interactions and higher-order synergistic associations~\cite{Quax2017} into the projection and replicating the analysis across diverse datasets, including more omics data such as genomics and proteomics.

\section*{Conclusions}

In this study, we introduced a novel multipartite projection method to construct multilayer disease networks, focusing on the comorbidity between CVD and depression. This approach complements traditional biological and experimental analyses by offering insights into the underlying biological pathways. By leveraging the MI correlation, our method integrates multiple omics datasets, such as metabolomics and lipidomics, to provide a system-level overview of the biological pathways contributing to this comorbidity.

Our analysis, applied to data from the Young Finns Study, identified significant biomarkers that mediate the relationship between CVD and depression. These biomarkers include metabolites like creatinine, valine, and phospholipids in very large high-density lipoproteins (HDL), as well as lipids such as sphingomyelins (SM), phosphatidylcholines (PC), Phosphatidylglycerols (PG), triacylglycerols (TAG), and diacylglycerols (DAG). Additionally, we highlighted sex and body mass index (BMI) as important risk factors influencing the comorbidity.

The proposed method addresses several limitations of previous approaches by incorporating multiple omics layers, accounting for non-linear correlations, and providing a weighted measure of biomarker contributions. This allows for a more nuanced understanding of disease associations and potential pathways.

The findings of this study underscore the importance of a multidimensional approach in studying complex disease interactions. The multipartite projection method offers a powerful tool for uncovering the intricate biological pathways and mechanisms underlying comorbid conditions, paving the way for more targeted preventive and therapeutic strategies. Future research could expand this methodology to other disease pairs and incorporate additional omics datasets to further enhance our understanding of disease comorbidities.

\section*{Funding}
The author(s) declare financial support was received for the research, authorship, and/or publication of this article. Jie Li, Jos A. Bosch and Rick Quax were supported by the EU project To\_Aition, which has received funding from the European Union's Horizon 2020 research and innovation programme under grant agreement No 848146. Cillian Hourican and Rick Quax were supported by the Netherlands Organisation for Health Research and Development (ZonMw), Open Competition Grant 09120012010063. The Young Finns Study has been financially supported by the Academy of Finland: grants 356405, 322098, 286284, 134309 (Eye), 126925, 121584, 124282, 129378 (Salve), 117797 (Gendi), and 141071 (Skidi); the Social Insurance Institution of Finland; Competitive State Research Financing of the Expert Responsibility area of Kuopio, Tampere and Turku University Hospitals (grant X51001); Juho Vainio Foundation; Paavo Nurmi Foundation; Finnish Foundation for Cardiovascular Research; Finnish Cultural Foundation; The Sigrid Juselius Foundation; Tampere Tuberculosis Foundation; Emil Aaltonen Foundation; Yrjö Jahnsson Foundation; Signe and Ane Gyllenberg Foundation; Diabetes Research Foundation of Finnish Diabetes Association; EU Horizon 2020 (grant 755320 for TAXINOMISIS and grant 848146 for To Aition); European Research Council (grant 742927 for MULTIEPIGEN project); Tampere University Hospital Supporting Foundation; Finnish Society of Clinical Chemistry; the Cancer Foundation Finland; pBETTER4U\_EU (Preventing obesity through Biologically and bEhaviorally Tailored inTERventions for you; project number: 101080117); and the Jane and Aatos Erkko Foundation. PPM was supported by the Academy of Finland (Grant number: 349708); M.A.-K. was supported by a research grant from the Sigrid Juselius Foundation, the Finnish Foundation for Cardiovascular Research, and the Research Council of Finland (grant no. 357183).

\bibliography{sample}



\section*{Author contributions statement}

JL wrote the first draft, performed all calculations, created the figures. JB and RQ conceptualized the study, provided feedback and supervision. CH participated in discussions, reviewed the manuscript, and provided feedback. Coauthors from the Young Finns Study (YFS) team facilitated access to the data. All authors have read and approved the final manuscript.

\section*{Conflicts of interest}

The authors declare no conflict of interest.

\section*{Data availability}
Data from the Young Finns Study can be accessed by contacting the study's data management team and complying with their access policies (http://youngfinnsstudy.utu.fi/).

\section*{Code availability}
The GitHub repository can be found at \url{https://github.com/ComplexNetSystem/multipartite_projection.git}.




\clearpage
\newpage

\makeatletter 
\renewcommand{\thetable}{\@arabic\c@table} 
\begin{table}
\advance\leftskip-0.5cm
\begin{center}
\begin{tabular}{cc}
\hline
\textbf{Variable codes} & \textbf{CVD-related phenotypes}  \\
\hline

dkv           & Diastolic KV blood pressure average, from the status form; mmHg     \\[1pt]
syst          & Systolic blood pressure average, from the status form; mmHg    \\[1pt]
fmd40         & Change in diameter 40s-base; mm    \\[1pt]
fmd40pr       & Change in diameter 40s-base in percentages; \%     \\[1pt]
fmd60         & Change in diameter 60s-base; mm     \\[1pt]
fmd60pr       & Change in diameter 60s-base in percentages; \%     \\[1pt]
fmd80         & Change in diameter 80s-base; mm     \\[1pt]
fmd80pr       & Change in diameter 80s-base in percentages; \%     \\[1pt]
maxfmd        & Maximum change in diameter; mm   \\[1pt]
maxfmdpr      & Maximum change in diameter in percentages; \%     \\[1pt]
pp            & Pulse pressure     \\[1pt]
idealCVH      & ideal cardiovascular health score 2007     \\[1pt]
imtka         & Carotid IMT average; mm     \\[1pt]
bbka          & Bulbus IMT average; mm     \\[1pt]
imtmax        & Carotid IMT maximum; mm    \\[1pt]
bbmax         & Bulbus IMT maximum; mm     \\[1pt]
volscore      & Total calcium score in volume of all coronary arteries; $mm^3$     \\[1pt]

\hline
\end{tabular}\caption{CVD-related phenotypes.}
\label{tab:1}
\end{center}
\end{table}

\makeatletter 
\renewcommand{\thetable}{\@arabic\c@table} 
\begin{table}
\advance\leftskip-0.5cm
\begin{center}
\begin{tabular}{cc}
\hline
\textbf{Variable codes} & \textbf{Depressive symptoms}  \\
\hline

b1                  & Sadness     \\[1pt]
b2                  & Pessimism    \\[1pt]
b3                  & Past failure    \\[1pt]
b4                  & Loss of pleasure     \\[1pt]
b5                  & Guilty feelings     \\[1pt]
b6                  & Punishment feelings     \\[1pt]
b7                  & Self-dislike     \\[1pt]
b8                  & Self-criticalness     \\[1pt]
b9                  & Suicidal thought or wishes    \\[1pt]
b10                 & Crying     \\[1pt]
b11                 & Agitation     \\[1pt]
b12                 & Loss of interest     \\[1pt]
b13                 & Indecisiveness     \\[1pt]
b14                 & Worthlessness     \\[1pt]
b15                 & Loss of energy     \\[1pt]
b16                 & Changes in sleep pattern     \\[1pt]
b17                 & Irritability     \\[1pt]
b18                 & Changes in appetite    \\[1pt]
b19                 & Concentration difficulty     \\[1pt]
b20                 & Tiredness or fatigue    \\[1pt]
b21                 & Loss of interest in sex     \\[1pt]
beckpisteet         & Beck Depression Inventory (BDI) score     \\[1pt]

\hline
\end{tabular}\caption{Depressive symptoms.}
\label{tab:2}
\end{center}
\end{table}

\makeatletter 
\renewcommand{\thetable}{\@arabic\c@table} 
\begin{table}
\advance\leftskip-0.5cm
\begin{center}
\begin{tabular}{ccc} 
\hline
\textbf{Omics} & \textbf{Variable code} & \textbf{Variable name} \\
\hline
\multirow{10}{5em}{NMR}         & Crea             & Creatinine; mmol/l                   \\ 
                                & XLHDLPL          & Phospholipids in very large HDL; mmol/l      \\ 
                                & SLDLTG           & Triglycerides in small LDL; mmol/l       \\ 
                                & Val              & Valine; mmol/l               \\ 
                                & ApoBApoA1        & Ratio of apolipoprotein B to apolipoprotein A-I       \\ 
                                & SVLDLPL          & Phospholipids in small VLDL; mmol/l                 \\ 
                                & LHDLFC           & Free cholesterol in large HDL; mmol/l               \\ 
                                & XSVLDLTG         & Triglycerides in very small VLDL; mmol/l             \\ 
                                & LVLDLFCPCT       & Free cholesterol to total lipids ratio in large VLDL; \%      \\ 
                                & Leu              & Leucine; mmol/l            \\ 
\hline
\multirow{10}{5em}{Lipids}      & SM.40.0                & Sphingomyelins 40:0  \\ 
                                & PC.38.4b               & Phosphatidylcholines 38:4b \\ 
                                & PC.36.4b..2            & Phosphatidylcholines 36:4b +2 \\ 
                                & TAG.16.0.18.0.18.1...1 & Triacylglycerols 16:0/18:0/18:1 +1 \\ 
                                & DAG.18.0.18.1.         & Diacylglycerols 18:0/18:1 \\ 
                                & TAG.18.0.18.1.18.1...1 & Triacylglycerols 18:0/18:1/18:1 +1 \\ 
                                & DAG.16.0.20.4.         & Diacylglycerols 16:0/20:4 \\ 
                                & SM.32.2                & Sphingomyelins 32:2 \\ 
                                & PG.36.1                & Phosphatidylglycerols 36:1 \\ 
                                & PE.32.1                & Phosphatidylethanolamines 32:1 \\                        
\hline
\end{tabular}\caption{Identified biomarkers related to CVD-depression comorbidity considering mean total contribution score.}
\label{tab:ttl_imp}
\end{center}
\end{table}

\makeatletter 
\renewcommand{\thetable}{\@arabic\c@table} 
\begin{table}
\advance\leftskip-0.5cm
\begin{center}
\begin{tabular}{ccc}
\hline
\textbf{Risk factors} & \textbf{CVD(\%)} & \textbf{Depression(\%)} \\
\hline
Age (ika07)                             & 4.25   & 6.39   \\[1pt]
Sex (SP)                                & 31.15  & 27.63  \\[1pt]
Socio-eco position (SESKOU07)           & 8.03   & 5.04   \\[1pt]
Smoke (smoke07)                         & 8.76   & 8.71   \\[1pt]
BMI (bmi07)                             & 40.82  & 47.64  \\[1pt]
Exercise (met07)                        & 7.00   & 4.58   \\[1pt]
\hline
\end{tabular}\caption{Relative importance of risk factors.}
\label{tab:importance}
\end{center}
\end{table}

\end{document}